\documentclass[prl,twocolumn,preprintnumbers,amsmath,amssymb,nofootinbib,showpacs]{revtex4}
\usepackage{graphicx}
\usepackage{amssymb}
\usepackage{mathrsfs}
\newcommand{\beq}{\begin{equation}}
\newcommand{\eeq}{\end{equation}}

\begin{document}

\title{On a scale-invariant Fermi gas in a time-dependent harmonic potential}
\date{\today}
\author{Sergej Moroz}

\affiliation{
\it Department of Physics, University of Washington \\
Seattle, WA 98195-1560, USA\\
}

\begin{abstract}
We investigate a scale-invariant two-component Fermi gas in a time-dependent isotropic harmonic potential. The exact time evolution of the density distribution in position space in any spatial dimension is obtained. Two experimentally relevant examples, an abrupt change and a periodic modulation of the trapping frequency are solved. Small deviations from scale invariance and isotropy of the confinement are addressed within first order perturbation theory. We discuss the consequences for experiments with ultracold quantum gases such as the excitation of a tower of undamped breathing modes and a new alternative for measuring the Tan contact.
\end{abstract}


\maketitle

\textit{Introduction:} Symmetries play a key role in modern physics. They can provide useful insights into understanding of systems whose microscopic dynamics is not known or poorly understood. If the microscopic description is available, symmetries serve as a guiding principle for the construction of solutions. In this work we study consequences of scale invariance which implies that physical observables do not depend on absolute lengths. We consider a two-component Fermi gas with contact interactions in any spatial dimension $d$ governed by the Hamiltonian \cite{hbar}
\beq \label{H}
H=\int d {\bf x} \Big{[} -\sum_{i=\uparrow,\downarrow}\psi_i^{\dagger}\frac{\nabla^2}{2m}\psi_i+c\psi^{\dagger}_\downarrow \psi^{\dagger}_\uparrow\psi_\uparrow\psi_\downarrow \Big{]}
\eeq
tuned to a scale-invariant regime and trapped in a time-dependent isotropic harmonic potential. The symmetries of this problem allow to find an exact time evolution provided the initial state is known. No knowledge of the equation of state or spectral functions is needed. This is especially beneficial for the much studied Fermi gas with infinite scattering length in $d=3$ which is theoretically one of the most interesting scale-invariant system. Symmetries alone predict a number of robust dynamical phenomena, which we illustrate here using two examples. The problems investigated in this paper can be directly realized in experiments with ultracold Fermi atoms.

\textit{Exact time evolution:}
Consider a Fermi gas loaded in a time-dependent isotropic harmonic potential described by the Hamiltonian
\beq \label{Hosc}
H_{\text{osc}}=H+\int d {\bf x}\frac{m \omega^2(t) {\bf x}^2}{2}\sum_{i=\uparrow,\downarrow}\psi_i^{\dagger}\psi_i.
\eeq
In the following, we assume that for $t<0$ the trap is static, i.e. $\omega(t<0)=\omega_{\text{in}}$, and that at $t=0$ the given $N$-body system  is in the eigenstate $\psi({\bf X})$ of the Hamiltonian $H_{\text{osc}}$ with the energy $E$. Here ${\bf X}$ collectively denotes the set of positions $({\bf x}_1,\dots, {\bf x}_N)$ of $N$ Fermi particles \cite{polarization}. Subsequently, for $t>0$ the trap frequency is varied with an arbitrary time dependence $\omega(t)$. The time evolution of a scale-invariant Fermi gas in $d$ spatial dimensions can be obtained from the initial wave-function by a combined gauge and scale transformation
\beq \label{ansatz}
\psi({\bf X}, t)=\frac{e^{-i\theta(t)}}{\lambda^{d N/2}(t)}\exp\left[\frac{i m \dot\lambda(t)}{2\lambda} X^2 \right]\psi({\bf X}/\lambda(t)),
\eeq
where $\dot\lambda(t)\equiv\frac{d\lambda(t)}{dt}$. Both $\theta(t)$ and $\lambda(t)$ are determined by the shape of $\omega(t)$. For one particle Eq. \eqref{ansatz} goes back to works \cite{LPP} that is easily generalized to any number $N$ of noninteracting particles \cite{Bruun}. Recently Castin made an insightful observation that the solution \eqref{ansatz} is also valid for the three-dimensional strongly-coupled unitary Fermi gas \cite{Castin2004}. This was achieved by showing that Eq. \eqref{ansatz} obeys the Bethe-Peierls contact condition at unitarity. We find that the solution \eqref{ansatz} is valid for a scale-invariant Fermi gas in any spatial dimension \cite{SM}.

One can check that for $\psi({\bf X},t)$ to be a solution, the gauge angle $\theta(t)$ must solve
$\dot \theta(t)=\frac{E}{\lambda^2(t)} \quad \text{with} \quad \theta(0)=0$,
while the scaling function $\lambda(t)$ obeys the differential equation
\beq \label{3}
\ddot{\lambda}(t)=\frac{\omega_{\text{in}}^2}{\lambda^3(t)}-\omega^2(t)\lambda(t), \qquad \omega_{\text{in}}\equiv \omega(t=0_{-})
\eeq
with the initial conditions
\beq \label{4}
\lambda(0)=1, \qquad \dot{\lambda}(0)=0.
\eeq
We recognize a one-dimensional Newton equation for a particle in an inverse square and a time-dependent harmonic potential. Physically, the scaling function $\lambda(t)$ is of a great interest, since it governs the time evolution of various observables. Among them the most experimentally relevant is the density distribution in position space that evolves as
$n({\bf x}, t)=\frac{1}{\lambda^d (t)} n_{0}(\frac{{\bf x}}{\lambda(t)})$,
where $n_{0}({\bf x})$ is an initial density profile at $t=0$. For the cloud of initial radius $r_{\text{cl}, 0}$ this implies
$r_{\text{cl}}(t)=\lambda(t)r_{\text{cl}, 0}$.
Since $\lambda(t)$ does not depend on energy $E$,  the latter two formulae are valid not only for pure states, but also hold for any initial statistical mixture of stationary states such as, for example, a thermal state.

\textit{Abrupt perturbation:}
First, we consider an experimental setting, where the frequency is changed abruptly at $t=0$ from the initial positive value $\omega_{\text{in}}$ to the final positive value $\omega_{\text{f}}$
\beq \label{freqevol}
\omega(t)=\omega_{\text{in}}+(\omega_{\text{f}}-\omega_{\text{in}})\theta(t).
\eeq
For $t>0$ the potential in the Newton equation \eqref{3} has a minimum at $\lambda_{\text{min}}=\sqrt{\frac{\omega_{\text{in}}}{\omega_{\text{f}}}}$ around which $\lambda(t)$ oscillates periodically starting from its initial state \eqref{4}.
In this case the exact solution of Eqs. \eqref{3} and \eqref{4} for $t>0$ can be found. Most easily the solution is obtained by introducing a new variable $\mu(t)\equiv\lambda^2(t)$. In terms of this variable, Eq. \eqref{3} transforms into a linear differential equation
$\ddot \mu(t)=2(\omega_{\text{in}}^2+\omega_{\text{f}}^2)-(2\omega_{\text{f}})^2 \mu(t)$
with the initial conditions $\mu(0)=1$ and $\dot\mu(0)=0$. This problem of a particle in the harmonic potential of frequency $2\omega_{\text{f}}$ subject to a constant force is easily solved and one obtains \cite{note}
\beq \label{6}
\begin{split}
\lambda(t)&=\sqrt{\frac{(1+\alpha)+(1-\alpha)\cos(2\omega_{\text{f}} t)}{2}} \\
          &=\sqrt{\cos^2(\omega_{\text{f}} t)+\alpha \sin^2(\omega_{\text{f}} t)},
\end{split}
\eeq
where $\alpha\equiv \left(\frac{\omega_{\text{in}}}{\omega_\text{f}} \right)^2$.
In the limit $\omega_{\text{in}}\ll \omega_{\text{f}}$ the inverse cube force in Eq. \eqref{3} is negligible for $\lambda>0$ and  acts only as an elastic boundary condition at $\lambda=0$. On the other hand, in the limit $\omega_{\text{in}}\gg \omega_{\text{f}}$ one finds the ballistic expansion of the cloud $\lambda(t)=\sqrt{1+\omega_{\text{in}}^2 t^2}$ for $t\ll\frac{\pi}{\omega_{\text{f}}}$.
The solution \eqref{6} can be expressed as the Fourier series
$\lambda(t)=\sum_{n=0}^{\infty}a_n \cos(2n \omega_{\text{f}} t)$
which physically corresponds to a decomposition into undamped isotropic breathing modes with frequencies $\omega_{n}=2n\omega_{\text{f}}$ and amplitudes $a_n$ \cite{linearized}.
 
\begin{figure}[t]
\begin{center}
\includegraphics[width=3.0in]{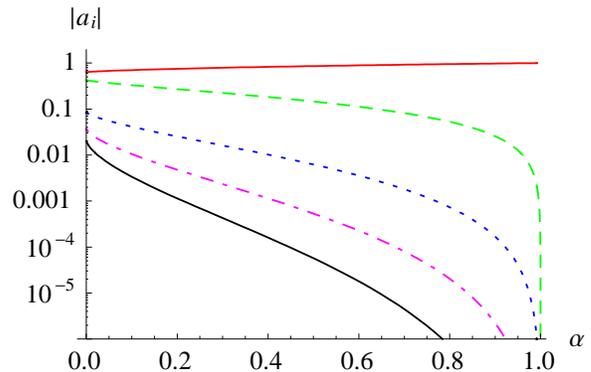}
\end{center}
\vskip -0.7cm \caption{Absolute values of the amplitudes $a_0$ (solid red), $a_1$ (dashed green), $a_2$ (dotted blue), $a_3$ (dashed-dotted magenta), $a_4$ (solid black) as a function of $\alpha=(\omega_{\text{in}}/\omega_{\text{f}})^2$.}
\label{fig1}
\end{figure}

For the step-function frequency profile \eqref{freqevol}, the lowest amplitudes $a_n$ can be computed analytically as a function of $\alpha$. Their absolute values are plotted in Fig. \ref{fig1} for $\alpha<1$ \cite{aboveone}.
Quite intuitively, the higher modes have smaller amplitudes compared to the lower ones. One can excite the higher modes most efficiently by a strong perturbation with $\omega_{\text{f}}\gg \omega_{\text{in}}$. In this limit we find $|a_0|=\frac{2}{\pi}$ and $|a_n|=\frac{4}{\pi (4n^2-1)}$ for $n\in\mathbf{N}$.

While to our knowledge a three-dimensional Fermi gas in a time-dependent isotropic trap has not yet been studied experimentally due to technical challenges, the two-dimensional case was recently investigated at different values of the scattering length $a_{\text{2d}}$ in \cite{Vogt}. In this experiment the collective breathing excitations were created by adiabatic reduction of the strength of the trapping frequency $\omega_{\perp}$ followed by an abrupt restoration to its original value. Provided the first adiabatic step does not excite collective modes, this experimental setting can be well described by Eq. \eqref{freqevol}. In \cite{Vogt} two different perturbations were studied: a weak perturbation with $\alpha=0.64$ and a strong one with $\alpha=0.36$. In both cases only the lowest breathing mode $\omega_1=2\omega_{\perp}$ was measured and no signature of the higher ones was detected. In the regime of asymptotic scale invariance \cite{away}, i.e. in the limit $a_{\text{2d}}\to\infty$ or $a_{\text{2d}}\to 0$, this fact can be understood from our calculation (see Fig. \ref{fig1}) which predicts $|\frac{a_2}{a_1}|\approx 3\%$ for the weak perturbation and $|\frac{a_2}{a_1}|\approx 6\%$ for the strong one. These are significantly below the experimental resolution limit $|\frac{a_2}{a_1}|\approx 20\%$ of the experiment \cite{Koehlprivate}. In future the higher breathing modes can be directly measured either by increasing the resolution limit of experiments or by enhancing the perturbation of the trap to values $\alpha\approx 0$.  

\textit{Periodic perturbation:}
Second, we investigate another experimentally relevant setting, where the trapping frequency oscillates periodically around its initial value $\omega_{\text{in}}$ as
\beq \label{perprofile}
\omega^2(t)=\omega^2_{\text{in}}+\Delta\omega^2 f(t).
\eeq
with $f(t+T)=f(t)$ \cite{remark}.
As the frequency varies in time, the initially stable equilibrium position $\lambda(0)=1$ of Eq. \eqref{3} can become unstable as more and more energy is pumped in. In the following we will first identify the condition for instability for a small perturbation with $0<\left(\Delta\omega/\Omega\right)^2\ll 1$, where $\Omega=2\pi/T$. To this end we notice that the solution of the nonlinear equation \eqref{3} with the initial conditions \eqref{4} can be related to the solution of the linear Newton equation for a time-dependent harmonic oscillator (with the same initial conditions)
\beq \label{linear}
\ddot{\gamma}(t)=-\omega^2(t)\gamma(t)
\eeq
via the formula
\beq \label{mapping}
\lambda^2(t)=\gamma^2(t)\left[1+\omega^2_{\text{in}}\xi^2(t) \right]
\eeq
with $\xi(t)=\int_{0}^{t}\frac{d\tau}{\gamma^2(\tau)}$ \cite{SM}. For $\omega^2_{\text{in}}>0$, due to the dominance of the repulsive inverse cube force near $\lambda=0$ in Eq. \eqref{3}, the scaling function $\lambda$ stays positive and finite at all times. For this reason $\xi\to\infty$ when $\gamma\to 0$. 
Time evolution beyond this point can be found by shifting $\xi\to -\infty$.

The stability analysis of the time-dependent harmonic oscillator \eqref{linear} is a textbook problem \cite{SM}. As a result, the instability known as a parametric resonance occurs if one period $T$ of the frequency modulation contains approximately a whole number of half-periods of the characteristic oscillations. In other words, for the resonant modulation frequencies we obtain
\beq \label{rescond}
\Omega_{n}=\frac{2\omega_{\text{in}}}{n}, \quad n\in \mathbf{N}.
\eeq
Since the inverse cube force in Eq. \eqref{3} is time-independent, it can not produce any additional resonances for $\lambda(t)$ in Eq. \eqref{mapping}. Hence the relation \eqref{rescond} is also a necessary and sufficient condition for the parametric resonance in the original nonlinear problem \eqref{3}. Therefore we arrive at a conclusion that there is an infinite set of modulation frequencies $\Omega_{n}$ \cite{linearized2} that will cause the atomic cloud of a scale-invariant Fermi gas to oscillate with the ever increasing amplitude up until energies where the zero-range description \eqref{H} breaks down. Actually, for a finite but small perturbation the resonant condition \eqref{rescond} must be satisfied only approximately. Generically, around every $\Omega_n$ there is a small band of resonant frequencies which scales as $\sim \Delta\omega \phi^n$, where $\phi\in(0,1)$ depends on the particular choice of the periodic function $f(t)$ \cite{Arnold}. Specifically, for the experimentally relevant $f(t)= \cos(\Omega t)$ the bands shrink as $\sim \Delta \omega^n$ \cite{Landau}. Thus in practice the resonances with smaller $n$ should be easier to observe.

\begin{figure}[t]
\begin{center}
\includegraphics[width=3.3in]{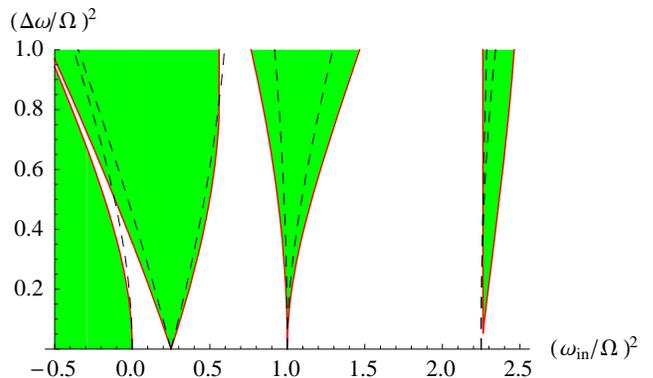}
\end{center}
\vskip -0.7cm \caption{Stability diagram parametrized by the dimensionless perturbation strength versus the dimensionless initial frequency. For $f(t)$ given by Eq. \eqref{periodicsteps} the solutions 
are unstable in the green-shaded region. Black dashed curves illustrate the instability boundary for $f(t)=\cos(\Omega t)$ \cite{Mathieu}.}
\label{fig2}
\end{figure}

What happens if the periodic perturbation is not small? In order to make quantitative predictions in this regime, we must specify the modulation function $f(t)$ in Eq. \eqref{perprofile}. For simplicity we take
\beq \label{periodicsteps}
f(t)=\left\{ \begin{array}{c} +1, \qquad t\in(0, T/2),   \\ -1, \qquad t\in(T/2, T). \\  \end{array} \right.
\eeq
As before, due to the mapping \eqref{mapping}, it is sufficient to analyze the stability of the time-dependent harmonic oscillator described by Eq. \eqref{linear}. The resonance condition now reads \cite{SM}
\beq \label{resonancefinite}
\begin{split}
2=&\Big{|}2\cos\left(\frac{\omega_+ T}{2}\right)\cos\left(\frac{\omega_- T}{2}\right)- \\
&\left(\frac{\omega_-}{\omega_+}+\frac{\omega_+}{\omega_-} \right)\sin\left(\frac{\omega_+ T}{2}\right)\sin\left(\frac{\omega_- T}{2}\right)\Big{|},
\end{split}
\eeq
where $\omega_{\pm}=\sqrt{\omega_{\text{in}}^2\pm \Delta\omega^2}$. The solution of this transcendental equation can be found numerically and is plotted in solid red in Fig. \ref{fig2}. For a weak perturbation with $\left(\Delta\omega/\Omega \right)^2\ll 1$ we recover the result \eqref{rescond}. As the strength of the perturbation increases the instability regions become broader. A notable feature of Fig. \ref{fig2} is that even the antitrapped Fermi gas with $\left(\omega_{\text{in}}/\Omega \right)^2<0$ can be stabilized by the properly tuned periodic perturbation \cite{negative}. This is a direct analogue of the inverted (Kapitza) pendulum stabilized by a vertically oscillating point of suspension.

\textit{Symmetries and beyond:}
In fact, the infinite tower of breathing modes in an isotropic trap is a general consequence of scale or more precisely of nonrelativistic conformal invariance \cite{Rosch, Castin, Nishida}. Indeed, using solely the generators ${\bf P}$, ${\bf K}$, $H$, $C$ and $D$  of the Schr\"odinger group (see \cite{Nishida} for the Schr\"odinger algebra and the definitions of these generators for the Fermi gas) we can construct the operators
\beq \label{generators} 
\begin{split} 
{\bf Q}^{\dagger}&=\frac{1}{\sqrt{2}}\left(\frac{ \bf P}{\sqrt{\omega}}+i \sqrt{\omega} {\bf K} \right),  \\  L^{\dagger}&=\frac{1}{2}\left(\frac{H}{\omega}-\omega C+iD \right).
\end{split}
\eeq
One can show that in the harmonic trap with the frequency $\omega$, the operator ${\bf Q}^{\dagger}$ excites center-of-mass energy eigenstates by acting repeatedly on a $N$-body primary state. Since
$[H_{\text{osc}}, {\bf Q}^{\dagger}]=\omega {\bf Q}^{\dagger}$,
the excited states have energies $E_0+n \omega$, where $n\in \mathbf{N}$ and $E_0$ denotes the energy of the primary state. In a similar fashion $L^{\dagger}$ excites breathing eigenstates with energies $E_0+2n \omega$. This is true since $L$, $L^{\dagger}$ and $H_{\text{osc}}$ satisfy
$[L, L^{\dagger}]=\frac{H_{\text{osc}}}{\omega}$, $ [H_{\text{osc}}, L^{\dagger}]=2\omega L^{\dagger}$.
The primary state must be annihilated by ${\bf Q}$ and $L$. It is worth emphasizing that while for $N=1$ the unique primary state is the ground state of the total Hamiltonian, for $N\ge 2$ there is an infinite number of the primary states. The energy spectrum is thus organized in infinite ladders with a ladder built on top of every primary state. 
While the individual center-of-mass (breathing) states do not actually deform in time because they are the eigenstates of the total Hamiltonian, the time evolution of a linear combination of the states from a given ladder produces dipole (breathing) density oscillation decomposable into modes with frequencies $n\omega$ ($2n\omega$). A simple way how to coherently excite such a linear combination is to perform the abrupt quench \eqref{freqevol}. It is clear from the solution \eqref{6} that the states from different ladders do not mix under such a rapid change of the trapping frequency \cite{delta}.


Due to separability of the center-of-mass and internal motion in a harmonic trap, one can construct the operator
\beq \label{breathingcr}
B^{\dagger}=L^{\dagger}-\frac{{\bf Q}^{\dagger}\cdot {\bf Q}^{\dagger}}{2m N}
\eeq
which excites internal breathing eigenstates. Indeed, $B$, $B^{\dagger}$ and $H_{\text{osc}}$ satisfy
$[B, B^{\dagger}]=\frac{H_{\text{osc}}}{\omega}-\frac{\{ Q_i, Q_i^{\dagger} \}}{2mN}$, $ [H_{\text{osc}}, B^{\dagger}]=2\omega B^{\dagger}$
and $B$ and $B^{\dagger}$ act only on the internal degrees of freedom of the atomic cloud \cite{LvsBremark}.

As argued above, the infinite equidistant tower of internal breathing modes is a generic feature of a scale-invariant many-body system loaded into an isotropic harmonic trap. But what happens to these modes if the symmetries are realized only approximately? Within first order perturbation theory the correction to the energy of the $n^{\text{th}}$ internal breathing state caused by a small symmetry-breaking Hamiltonian perturbation $\delta H$ is
\beq \label{bs1}
\delta E_n=\frac{\langle 0 | B^n \delta H B^{\dagger n}|0\rangle}{\langle0| B^n B^{\dagger n}|0\rangle},
\eeq
where $|0\rangle$ stands for a $N$-body primary state in the harmonic trap and $B^{\dagger}$ is defined in Eq. \eqref{breathingcr}. Here we assume that the internal breathing states are non-degenerate with other energy eigenstates in the trap. This should be fulfilled in the strongly interacting unitary Fermi gas in three spatial dimensions which we restrict our attention to in the following.

Consider first the breaking of scale invariance by a finite (but large) scattering length $a_{3d}$. For the Fermi gas near the unitarity regime the Hamiltonian perturbation can be expressed using the local composite dimer field $\phi$ via
\beq \label{Hp}
\delta H=-\frac{m a_{3d}^{-1}}{4\pi} \int d {\bf x} \phi^{\dagger} \phi.
\eeq
Since this perturbation does not affect the motion of the center of mass, $\delta E_n$ equals to the energy shift $\delta \mathscr{E}_n$ associated with the internal motion only.
By substituting this perturbation into Eq. \eqref{bs1} and using general properties of nonrelativistic scale invariance, we derive \cite{SM} for the shift of the level spacing $\delta \Delta_n=\delta (\mathscr{E}_n-\mathscr{E}_{n-1})=\delta (\omega_n-\omega_{n-1})$
\beq \label{rec}
\delta\Delta_n=\frac{S_{n-2}}{S_{n-1}}\delta\Delta_{n-1}-\frac{\omega}{4S_{n-1}}\delta \mathscr{E}_{n-1},
\eeq
where $S_{k}=(k+1)(\mathscr{E}_0+k \omega)$ and $\delta \Delta_0=0$. Provided the internal part of the energy $\mathscr{E}_0=E_0-3\omega/2$ at unitarity and its shift $\delta \mathscr{E}_0$ are known for a given primary state of the $N$-particle system, the recursion relation predicts the frequency shifts of the whole tower of breathing modes. The deviations from the scale-invariant value $\Delta_n=2\omega$ are the largest for the lowest breathing modes. At high energies as $n\to \infty$ the shift $\delta \Delta_n\sim n^{-3/2}\to 0$ \cite{SM}.

A precise experimental measurement of the lowest level spacing shifts $\delta\Delta_n$ in a many-particle Fermi gas near unitarity can provide a new way to measure the Bertsch parameter $\xi_{\text{B}}$ and the Tan contact $C_{\text{trap}}$. Indeed, at $T=0$ the local density approximation predicts for the ground state energy $\mathscr{E}_0\approx E_0=3^{4/3}\sqrt{\xi_{\text{B}}}N^{4/3}\omega/8$ \cite{rev}. On the other hand, the contact can be directly extracted from the energy shift via $\delta\mathscr{E}_0=-m a^{-1}_{\text{3d}}C_{\text{trap}}/4\pi$ \cite{Tan}. By substituting these two expressions into Eq. \eqref{rec} we obtain for the lowest shift
\beq
\delta\Delta_1=\frac{m C_{\text{trap}} a^{-1}_{\text{3d}}}{2\cdot 3^{4/3}\pi \sqrt{\xi_{\text{B}}}N^{4/3}}.
\eeq  
which allows to determine the ratio $C_{\text{trap}}/\sqrt{\xi_{\text{B}}}$. An additional measurement of $\delta\Delta_2$ would allow to extract separate values of $\xi_{\text{B}}$ and $C_{\text{trap}}$ from Eq. \eqref{rec}.

In experiments scale invariance is broken by a finite effective range $r_{\text{eff}}$. In this case the Hamiltonian perturbation can be expressed as
\beq 
\begin{split} \label{range}
 \delta H&= \frac{m r_{\text{eff}}}{16 \pi} \int d {\bf x} \left( \phi^{\dagger}[H_{\text{int}}, \phi] +\text{c. c.} \right) \\
&=\frac{m r_{\text{eff}}}{16 \pi} \int d {\bf x} \left( \phi^{\dagger}(-i\partial_t-H_{\text{CM}}) \phi +\text{c. c.} \right),
\end{split}
\eeq
where $H_{\text{int}}=H_{\text{osc}}-H_{\text{CM}}$ with $H_{\text{CM}}=\omega \{Q_i, Q_i^{\dagger} \}/2mN$. This perturbation is invariant under translations and Galilean boosts, in addition it affects only the internal motion. For the perturbation \eqref{range} we find the recursion relation \cite{SM}
\beq \label{rec1}
\delta\Delta_n=\frac{S_{n-2}}{S_{n-1}}\delta\Delta_{n-1}+\frac{3\omega}{4S_{n-1}}\delta \mathscr{E}_{n-1}.
\eeq
For high levels the shift scales as $\delta\Delta_n\sim n^{-1/2}\to 0$ when $n\to\infty$ \cite{SM}. We found that the relations $\eqref{rec}$ and $\eqref{rec1}$ are in agreement with the perturbation expansion around unitarity, done recently in \cite{Daily}, of the analytical solution for two particles in a harmonic trap \cite{Busch}.

Let us also consider a long-range two-body (an)isotropic perturbation of the form
\beq \label{lrange}
\delta H\sim \int d {\bf x} d {\bf y} n({\bf x}) \frac{g(\theta)}{|{\bf x}-{\bf y}|^{\rho}} n({\bf y}) 
\eeq
with $n=\sum_{i=\uparrow,\downarrow}\psi_i^{\dagger}\psi_i$, $\rho\in \mathbf{R}$ and $g(\theta)$ is some function of the angle between the unit vector pointing at some fixed direction (e.g. induced by an external field) and the vector ${\bf x}-{\bf y}$. The recursion relation can be readily found to be \cite{SM}
\beq \label{rec2}
\delta\Delta_n=\frac{S_{n-2}}{S_{n-1}}\delta\Delta_{n-1}+\frac{\alpha(\alpha-2)\omega}{4S_{n-1}}\delta \mathscr{E}_{n-1}.
\eeq
Note that for the inverse-square interaction potential ($\rho=2$) the perturbation is scale-invariant and thus does not modify the breathing frequencies. In the context of cold atom experiments, Eq. \eqref{rec2} allows to estimate the effect of weak magnetic dipole-dipole interactions ($\rho=3$) on the tower of breathing frequencies.

Finally, we investigate a small deviation from isotropy of the harmonic confinement originating from the perturbation
\beq \label{aniso}
\delta H= \frac{m \delta \omega^2}{2} \int d {\bf x} \, x_{\parallel}^2\sum_{i=\uparrow,\downarrow}\psi_i^{\dagger}\psi_i,
\eeq
where $x_{\parallel}$ is the set of coordinates with $\omega^2\to \omega^2+\delta \omega^2$ imposed.
First order perturbation theory gives a very simple prediction for the shifts of the level spacing \cite{SM}. Namely,
\beq \label{aniso1}
\delta \Delta_n=2\omega \frac{\delta \mathscr{E}_0}{\mathscr{E}_0} \quad \text{for} \quad n\in\mathbf{N}
\eeq 
meaning that the whole tower of breathing frequencies is homogeneously stretched by a factor $\left(1+\delta\mathscr{E}_0/\mathscr{E}_0 \right)$.
We checked that for a small anisotropy Eq. \eqref{aniso1} agrees with the analytic solution for two particles in an anisotropic trap \cite{Idziaszek}.

\textit{Conclusion:}
In this work we studied a scale-invariant Fermi gas in a time-dependent isotropic harmonic potential. Within the zero-range model \eqref{H} the exact time evolution can be found by solving an effective one-dimensional Newton equation. As examples we considered two experimentally relevant settings. First, an abrupt change of the trapping frequency $\omega(t)$ in the form of a step function was studied. We found the exact solution of this problem, decomposed it into a series of breathing modes and discussed why only the lowest mode was observed in the recent experiment \cite{Vogt}. The influence of a small deviation from scale invariance and isotropy of the harmonic confinement on the frequencies of breathing modes was studied using first order perturbation theory.  Second, periodic oscillations around the initial value of the trapping frequency were investigated. We identified modulation frequencies at which the system becomes unstable and exhibits parametric resonances. We also observed that an antitrapped Fermi gas can be stabilized by periodic frequency oscillations.  The findings of this paper are valid at arbitrary temperature provided the zero-range model \eqref{H} accurately describes the Fermi gas. Higher breathing modes, parametric resonances and the Kapitza pendulum investigated in this work can be directly realized in future experiments with ultracold quantum gases.

\textit{Acknowledgment:}
It is our pleasure to acknowledge discussions with I. Boettcher, A. Bulgac, Y. Castin, S. Gupta, M. Koehl, D. Morozova, R. Schmidt, D. T. Son and W. Zwerger. This work was supported by U.S. DOE Grant No. DE-FG02-97ER41014.


\clearpage

\begin{center}
{\bf Supplemental material}
\end{center}

\textit{Scale invariance and contact Bethe-Peierls condition:}
Due to the contact nature of the interaction term in Eq. \eqref{H}, one can reformulate the $N$-body quantum-mechanical problem as a noninteracting one with the effect of interactions incorporated in the boundary conditions for the many-body wave function \cite{Castinreview}. In particular in $d\ne 2$, when any spin-up fermion approaches any spin-down fermion the many-body wave function $\Psi$ must obey the Bethe-Peierls boundary condition
\beq \label{BP}
\Psi(r)\sim\frac{B_1}{r^{d-2}}+B_2 \quad \text{for} \quad r\to 0,
\eeq
where $r$ denotes the distance between the two fermions and we suppressed remaining arguments of the wave function. The condition \eqref{BP} can be most easily found by solving the two-body problem with a pseudopotential, but it is valid for any $N$. The generic boundary condition \eqref{BP} is not scale-invariant since the ratio $B_1/B_2$ carries a physical dimension. A scale-invariant Fermi gas is obtained by taking $B_1=0$ or $B_2=0$. For $d>4$ the solution with $B_1\ne 0$ is not normalizable and thus is forbidden by unitarity \cite{Nussinov}. Therefore, for $d>4$ only a noninteracting scale-invariant Fermi gas is allowed.

In order to illustrate the regimes of applicability of this paper, we will list all scale-invariant solutions in $d\le3$:
\begin{itemize}
\item $\mathbf{d=3.}$ The quantum-mechanical s-wave scattering of two fermions with the relative momentum $q$ leads to the amplitude \cite{reviews}
\beq \label{f3d}
f_{\text{3d}}(q)=\frac{1}{-a_{\text{3d}}^{-1}-i q}
\eeq
with the scattering length $a_{\text{3d}}$. By dimensional analysis $B_1/B_2\sim a_{\text{3d}}$. Fermions with $a_{\text{3d}}=0$ constitute a simple noninteracting scale-invariant system. In addition, scale invariance is also attained at the unitarity point $a^{-1}_{\text{3d}}=0$, where the theory is strongly coupled.
\item $\mathbf{d=2.}$ This case is special as the coupling strength $c$ in Eq. \eqref{H} is dimensionless. In $d=2$ the boundary condition $\eqref{BP}$ must be replaced by
\beq \label{BP2}
\Psi(r)\sim B_1 \ln r k+B_2
\eeq
with $k$ denoting an arbitrary momentum scale. The scattering amplitude now reads \cite{reviews}
\beq \label{f2d}
f_{\text{2d}}(q)=\frac{4\pi}{\ln\left(1/q^2a_{\text{2d}}^2\right)+i\pi},
\eeq
where $a_{\text{2d}}$ is a (positive) scattering length. For the interacting Fermi gas one gets $B_1/B_2\sim\ln \left( 1/a_{\text{2d}} k \right)$. Apart from the strictly noninteracting regime with $B_1/B_2=0$, scale invariance is approached asymptotically at low energies (or equivalently $a_{\text{2d}}\to 0$) in the repulsive Fermi gas or at high energies (or equivalently $a_{\text{2d}}\to +\infty$) in the attractive Fermi gas.
\item $\mathbf{d=1.}$ The scattering amplitude (or equivalently the reflective amplitude) is given by \cite{reviews, Barth}
\beq \label{f1d}
f_{\text{1d}}(q)=-\frac{1}{1+iq a_{\text{1d}}},
\eeq
which defines the scattering length $a_{\text{1d}}$. From dimensional arguments $B_1/B_2\sim a_{\text{1d}}^{-1}$. Note that no regularization is needed in this case and $c=-1/m a_{\text{1d}}$. Thus the noninteracting scale-invariant regime is approached as $a_{\text{1d}}^{-1}\to 0$. In addition, the interacting scale-invariant solutions are: an infinitely repulsive unitary Fermi gas with $c\to+\infty$, i.e. $a_{\text{1d}}\to 0_-$ (which is equivalent to a noninteracting one-component Fermi gas) and an infinitely attractive Fermi gas with $c\to-\infty$, i.e. $a_{\text{1d}}\to 0_+$ (which can be viewed as the Tonks--Girardeau limit of the Bose gas of compact dimers).
\end{itemize}

From the discussion above it becomes clear that apart from the unitary Fermi gas in $d=3$, any scale-invariant nonrelativistic Fermi gas is noninteracting or can be mapped onto a noninteracting system.

Finally, we demonstrate that the exact time evolution of any scale-invariant Fermi gas governed by the Hamiltonian \eqref{Hosc} is given by Eq. \eqref{ansatz}. To this end we first verified that Eq. \eqref{ansatz} solves the free many-body Schr\"odinger equation. In addition, it is necessary to show that the scale-invariant Bethe-Peierls boundary condition, which is a pure power law, is preserved by the time-evolution \eqref{ansatz}. This is indeed the case since the gauge factor
\beq
\exp\left[\frac{i m \dot\lambda(t)}{2\lambda} r^2 \right]=1+O(r^2)
\eeq
and thus does not modify the leading power law exponent.

\textit{Proof of Eq. (\ref{mapping}):}
To justify the formula $\eqref{mapping}$ we demonstrate how the solution of the nonlinear (Ermakov) differential equation \eqref{3} can be related to the solution of the linear Newton equation for a time-dependent harmonic oscillator \eqref{linear}. To this end, one introduces a new coordinate $\nu$ and time $\xi$ via
\beq \label{new}
\nu=\frac{\lambda}{\gamma}, \qquad \xi=\int_{0}^{t}\frac{d\tau}{\gamma^2(\tau)},
\eeq
which transforms the Ermakov equation to
\beq
\frac{d^2 \nu}{d \xi^2}=\frac{\omega_{\text{in}}^2}{\nu^3(\xi)}.
\eeq
This can now be easily integrated leading to the general solution
\beq
C_1 \lambda^2=\omega_{\text{in}}^2\gamma^2+\left(C_1\xi+C_2 \right)^2\gamma^2
\eeq
found first in \cite{Ermakov}.
By substituting the initial conditions \eqref{4} for $\lambda$ into this solution and applying the same initial conditions for $\gamma$ one gets $C_1=\omega^2_{\text{in}}$ and $C_2=0$ which lead directly to Eq. \eqref{mapping}.


\textit{Mapping at a period and stability condition:}
In order to identify the resonant frequencies of the harmonic oscillator driven by a weak periodic perturbation, we follow \cite{ArnoldS} and introduce a linear operator $A$ that evolves an arbitrary initial state in the phase space at a period $T$, i.e.
\beq
(\gamma(0),\dot\gamma(0))\xrightarrow{A} (\gamma(T),\dot\gamma(T)).
\eeq
For vanishing perturbation $\Delta\omega\to 0$ one gets
\beq \label{A}
A(\omega_{\text{in}})=\left(
\begin{array}{cc}
\cos 2\pi \frac{\omega_{\text{in}}}{\Omega} & \frac{1}{\omega_{\text{in}}}\sin 2\pi \frac{\omega_{\text{in}}}{\Omega}  \\ 
-\omega_{\text{in}}\sin 2\pi \frac{\omega_{\text{in}}}{\Omega} & \cos 2\pi \frac{\omega_{\text{in}}}{\Omega}
\end{array} \right).
\eeq
According to the stability theorem from \cite{ArnoldS}, a linear hamiltonian system is stable if $|\text{tr}A|>2$. Specifically, for Eq. \eqref{A} we find $|\text{tr} A(\omega_{\text{in}})|=2|\cos 2\pi \frac{\omega_{\text{in}}}{\Omega}|>2$ unless $\Omega=\Omega_n=2\omega_{\text{in}}/n$. Thus for $\Delta\omega\to 0$ the driven oscillator \eqref{linear} becomes unstable only at $\Omega=\Omega_n$ where the resonance phenomenon occurs. Note that for $\Delta\omega\to 0$ the anti-trapped (i.e. $\omega_{\text{in}}^2<0$) solution is always unstable.

For the finite periodic perturbation with the modulation function \eqref{periodicsteps} the linear operator $A$ can be expressed as
\beq
A=A\left(\frac{\omega_+}{2} \right) A\left(\frac{\omega_-}{2} \right)
\eeq
with $\omega_{\pm}=\sqrt{\omega_{\text{in}}^2\pm \Delta\omega^2}$. The resonance condition $|\text{tr}A|=2$ now leads to Eq. \eqref{resonancefinite}.

\textit{Shifts of breathing frequencies:}
We start from Eq. \eqref{bs1} and commute in the numerator one operator $B$ to the very right, where it annihilates the primary state $|0\rangle$, obtaining
\beq \label{fa}
\delta E_n=\frac{\langle n-1|[B,\delta H]B^{\dagger}|n-1\rangle}{\langle n| n \rangle}+\delta E_{n-1},
\eeq 
where $|n\rangle=B^{\dagger n}|0\rangle$. Here we used $[B,B^{\dagger n}]|0\rangle=\frac{1}{\omega} \sum_{i=0}^{n-1} \mathscr{E}_i |n-1\rangle$ and $\langle n|n\rangle=\prod_{k=1}^{n}{\frac{1}{\omega}\sum_{i_k=0}^{k-1}\mathscr{E}_{i_{k}}}$. Now by commuting one creation operator $B^{\dagger}$ to the very left in the numerator of the first term in Eq. \eqref{fa}, we find for $\delta \Delta_n=\delta E_n-\delta E_{n-1}=\delta \mathscr{E}_n-\delta \mathscr{E}_{n-1}$
\beq \label{d2}
\delta\Delta_n=\frac{S_{n-2}}{S_{n-1}}\delta\Delta_{n-1}+\frac{\omega}{S_{n-1}}\frac{\langle n-1|[[B,\delta H],B^{\dagger}]|n-1\rangle}{\langle n-1|n-1\rangle},
\eeq
where $S_k=\sum_{i=0}^{k} \mathscr{E}_i=\sum_{i=0}^{k}(\mathscr{E}_0+2i\omega)=(k+1)(\mathscr{E}_0+k\omega)$ was introduced. The main challenge now is to evaluate the numerator of the second term in Eq. \eqref{d2}. To this end we notice that the matrix element does not depend on time and for our convenience we evaluate it in the Heisenberg picture at $t=0$. In addition, for the perturbation which commutes with the center-of-mass operators ${\bf Q}^{\dagger}$ and ${\bf Q}$ (defined in Eq. \eqref{generators}) we can replace
\beq \label{sub}
[[B, \delta H],B^{\dagger}]\rightarrow [[L, \delta H],L^{\dagger}].
\eeq
Using $H=H_{\text{osc}}-\omega^2 C$, we find the inner commutator to be
\beq
\begin{split}
O\equiv[L, \delta H]&=\frac{1}{2}[\frac{1}{\omega}H_{\text{osc}}-2\omega C- i D, \delta H] \\
&=\frac{1}{2}\left(\frac{1}{\omega}[H_{\text{osc}}, \delta H]-2\omega [C,\delta H]+\Delta_{\delta H}\delta H \right),
\end{split}
\eeq
where $[D, \delta H]=i\Delta_{\delta H}\delta H$ (valid at $t=0$) was used \cite{NishidaS}. Here $\Delta_{\delta H}$ denotes the scaling dimension of the Hamiltonian perturbation which is given by
$
\Delta_{\delta H}=2\Delta_{\phi}-d
$
and
$
\Delta_{\delta H}=\rho
$
for the perturbation \eqref{Hp} and \eqref{lrange} respectively. On the other hand, the effective range perturbation \eqref{range} is a sum of two operators with different scaling dimensions $2\Delta_{\phi}-d+2$ and $2\Delta_{\phi}-d-2$.

To evaluate the outer commutator we note that $\langle k|[H_{\text{osc}},\text{anything}]|k\rangle=0$ for any energy eigenstate $|k\rangle$. Hence
\beq \label{outer}
\begin{split}
[O,L^{\dagger}]&=\frac{1}{2}\left[O,-2\omega C+iD \right] \\
&=\frac{1}{4}\Delta_{\delta H}(\Delta_{\delta H}-2)\delta H+\omega^2[[C,\delta H],C],
\end{split}
\eeq
where first the Jacobi identity was employed and the commutators
\beq
[C, H_{\text{osc}}]=i D, \quad [D, H_{\text{osc}}]=2i (H_{\text{osc}}-2\omega^2 C)
\eeq
followed from the Schr\"odinger algebra \cite{NishidaS}. 

Combining Eqs. \eqref{d2} and \eqref{outer}, one finally obtains
\beq \label{lsf0}
\begin{split}
\delta\Delta_n=\frac{S_{n-2}}{S_{n-1}}\delta\Delta_{n-1}&+\frac{\omega}{4S_{n-1}}\Delta_{\delta H}(\Delta_{\delta H}-2)\delta \mathscr{E}_{n-1} \\
&+\frac{\omega^3}{S_{n-1}}\frac{\langle n-1|[[C,\delta H],C^{\dagger}]|n-1\rangle}{\langle n-1|n-1\rangle}.
\end{split}
\eeq
Note that the last term in Eq. \eqref{lsf} does not contribute if $[\delta H, C]=0$ which is true for all perturbations discussed in this paper. Finally thus we obtain
\beq \label{lsf}
\delta\Delta_n=\frac{S_{n-2}}{S_{n-1}}\delta\Delta_{n-1}+\frac{\omega}{4S_{n-1}}\Delta_{\delta H}(\Delta_{\delta H}-2)\delta \mathscr{E}_{n-1}.
\eeq
In the unitary regime $\Delta_{\phi}=2$ in $d=3$ which leads to Eqs. \eqref{rec}, \eqref{rec1}  and \eqref{rec2} for the perturbations \eqref{Hp}, \eqref{range} and \eqref{lrange} respectively.

Consider now the level shifts caused by the anisotropic perturbation \eqref{aniso}. Since it affects both the center-of-mass and internal degrees of freedom, one can not perform the substitution \eqref{sub}. The calculation can be simplified, however, by noting that the perturbation can be replaced by the special conformal generator $C$ inside the double commutator in Eq. \eqref{d2}. Indeed,
\beq
\delta H \rightarrow \delta \omega^2 \frac{d_{\parallel}}{d} \underbrace{\frac{m}{2} \int d {\bf x} \, {\bf x}^2\sum_{i=\uparrow,\downarrow}\psi_i^{\dagger}\psi_i}_{C}.
\eeq
Here $d_{\parallel}$ is the number of dimensions affected by the anisotropic perturbation. It is a straightforward (but tedious) exercise in the Schr\"odinger algebra to evaluate the double commutator and obtain
\beq
\begin{split}
\frac{\langle n-1|[[B,\delta H],B^{\dagger}]|n-1\rangle}{\langle n-1|n-1\rangle}&=2\frac{\langle n-1|\delta H_{\text{int}}|n-1\rangle}{\langle n-1|n-1\rangle}\\
&=2(\delta E_{n-1}-\delta E_{\text{CM}, n-1})\\
&=2\delta\mathscr{E}_{n-1},
\end{split}
\eeq 
where $\delta H_{\text{int}}=\delta H -\delta \omega^2\frac{{\bf K}^2}{2m N}$ and ${\bf K}^2=m^2 N^2 {\bf X}^2_{\text{CM}}$ with ${\bf X}_{\text{CM}}$ denoting the center-of-mass position.
By substituting the last result into Eq. \eqref{d2}, we find that the level shift formula \eqref{lsf} is actually valid also for the anisotropic perturbation since now $\Delta_{\delta H}=-2$. As a result, the first level spacing is shifted by $\delta \Delta_1=2\omega \delta\mathscr{E}_0/\mathscr{E}_0$ and it is straightforward to show by induction that $\delta\Delta_n=\delta\Delta_1$ for all $n\in\mathbf{N}$.

\textit{Large $n$ asymptotics:}
Here we find the asymptotic solution of the recursion relation \eqref{lsf} for $n\to \infty$. To this end we consider a shallow harmonic potential with $\omega\to 0$ such that $x\equiv \omega (n-1)$ is finite for $n\to\infty$ and introduce a dimensionless energy shift $f(x)\equiv \delta\mathscr{E}_{n-1}/\omega$. In this limit for any finite number of particles $\mathscr{E}_0\ll x$ and thus $S_{n-2}/S_{n-1}\rightarrow 1-2\omega/x$ and $S_{n-1}\rightarrow x^2/\omega$. Using this we can cast the general recursion relation \eqref{lsf} into the continuous form of the second order differential equation
\beq
f^{''}=-\frac{2}{x}f^{'}+\frac{\alpha}{x^2}f,
\eeq
where $\alpha=\Delta_{\delta H}(\Delta_{\delta H}-2)/4$. The general solution of this equation is found to be
\beq \label{pl}
f(x)=C_+ x^{p_+}+C_- x^{p_-}
\eeq
with $p_{\pm}=\left(-1\pm \sqrt{1+4\alpha} \right)/2$. As $x\to\infty$ the second term in Eq. \eqref{pl} becomes small compared with the first one and can be neglected. In this way we find a pure power law asymptotics $\delta\mathscr{E}_n\sim f(x)\sim n^{p_+}$ and $\delta \Delta_n\sim f^{'}(x)\sim n^{p_+-1}$ for $n\to \infty$.


\begin{thebibliography}{11}
\bibitem{hbar}
$\hbar$ is set to unity in this paper, $m$ denotes the mass of fermions and $c$ stands for the coupling strength.
\bibitem{polarization}
 We put no restrictions on the spin polarization of the $N$-body Fermi system.
\bibitem{note}
Actually, the same solution was found previously in the context of evolution of a harmonically confined one-dimensional Tonks-Girardeau Bose gas \cite{Minguzzi}.
\bibitem{linearized}
Note that the linearized solution of Eq. \eqref{3} with the step-function profile $\omega(t)$ uncovers only the first breathing mode $\omega_{1}=2\omega_{\text{f}}$.
\bibitem{aboveone}
The solution \eqref{6} is valid also for $\alpha>1$. In this case the absolute values of the amplitudes $a_n$ can be found from the solution for $\alpha<1$ by using the symmetry relation $|a_n(\alpha^{-1})|=\frac{|a_n(\alpha)|}{\sqrt{\alpha}}$.
\bibitem{away}
Away from scale-invariance, i.e. at finite and non-vanishing $a_{\text{2d}}$, the breathing frequencies are predicted to be shifted by a quantum scale anomaly (see \cite{Hofmann} for the recent calculation of the shift of the lowest breathing mode).
\bibitem{linearized2}
Note that the linearization of Eq. \eqref{3} around the time-dependent minimum $\lambda_{\text{min}}(t)$ leads to $\ddot\sigma=1/4 \left(\Delta\omega/\omega_{\text{in}} \right)^2 \ddot f- 4\omega_{\text{in}}^2 \sigma$ for small oscillations $\sigma(t)=\lambda(t)-\lambda_{\text{min}}(t)$. This is just a harmonic oscillator with frequency $2\omega_{\text{in}}$ in a periodic external field which exhibits a resonance for $T=\pi/\omega$. Thus, the linearized solution unveils only one resonance frequency and misses completely the rest of the infinite tower.
\bibitem{Mathieu}
For $f(t)=\cos(\Omega t)$ the instability boundary for Eq. \eqref{linear}, known as the Mathieu equation, was found using the commands MathieuCharacteristic(A/B) of Mathematica.
\bibitem{negative}
For $\omega_{\text{in}}^2<0$, the factor $\left[1+\omega^2_{\text{in}}\xi^2(t) \right]$ in Eq. \eqref{mapping} must be replaced by its absolute value.
\bibitem{delta}
Another simple way how to excite breathing oscillations is a delta function perturbation $\omega^2(t)=\omega^2+\chi \delta(t)$ which leads to the solution $\lambda(t)=\sqrt{1-\chi \sin(2\omega t)/\omega+\chi^2 \sin^2(\omega t)/\omega^2}$.
\end{thebibliography}
\end{document}